\pdfoutput=1
\documentclass[letterpaper, 10 pt, conference]{ieeeconf}  

\IEEEoverridecommandlockouts                              
\usepackage{graphics}
\usepackage{epsfig} 
\usepackage{amsmath} 
\usepackage{amssymb}  

\newcommand {\matr}[2]{\left[\begin{array}{#1}#2\end{array}\right]}

\newcommand{\x}{{\mathbf{x}}}
\renewcommand{\u}{{\mathbf{u}}}

\renewcommand{\r}{{\mathbf{r}}}

\usepackage[usenames,dvipsnames]{color}
\definecolor{wheat}{rgb}{0.96,0.87,0.70}

\usepackage{tikz}
\usepackage{soul}

\title{\LARGE \bf
	Real-Time Constrained Trajectory Planning and Vehicle Control for Proactive Autonomous Driving With Road Users}

\author{Ivo Batkovic$^{1,2}$, Mario Zanon$^{3}$, Mohammad Ali$^{2}$, and Paolo Falcone$^{1}$
	\thanks{This work was partially supported by the Wallenberg Artificial Intelligence, Autonomous Systems and Software Program (WASP) funded by Knut and Alice Wallenberg Foundation, and by the COPPLAR project (VINNOVA. V.P. Grant No. 2015-04849).}	
	\thanks{$^{1}$ Ivo Batkovic and Paolo Falcone are with the Mechatronics group at the Department of Electrical Engineering, Chalmers University of Technology, Gothenburg, Sweden {\tt\footnotesize \{ivo.batkovic,falcone\}@chalmers.se }}%
	\thanks{$^{2}$ Ivo Batkovic, and Mohammad Ali are with the research department at Zenuity AB {\tt\footnotesize \{ivo.batkovic,mohammad.ali\}@zenuity.com}}%
	\thanks{$^{3}$ Mario Zanon is with the IMT School for Advanced Studies Lucca {\tt\footnotesize mario.zanon@imtlucca.it}}%
}

\begin{document}
	
	\maketitle
	\thispagestyle{empty}
	\pagestyle{empty}
	
	\begin{abstract}
		For motion planning and control of autonomous vehicles to be proactive and safe,  pedestrians' and other road users' motions must be considered. In this paper, we present a vehicle motion planning and control framework, based on Model Predictive Control, accounting for moving obstacles. Measured pedestrian states are fed into a prediction layer which translates each pedestrians' predicted motion into constraints for the MPC problem. 
		
		
		Simulations and experimental validation were performed with simulated crossing pedestrians to show the performance of the framework. Experimental results show that the controller is stable even under significant input delays, while still maintaining very low computational times. In addition, real pedestrian data was used to further validate the developed framework in simulations.

		
	\end{abstract}

	\section{Introduction}
	Since a decade, autonomous driving technologies have emerged with the potential for increased safe and efficient driving~\cite{ziegler2014making}. While one can expect autonomous driving technologies to be deployed first in structured environments such as highway driving and low-speed parking~\cite{becker2014bosch}, other scenarios, such as urban driving, arguably pose a greater challenge due to the presence of non-autonomous road users, e.g. pedestrians, cyclists, and other vehicles. Hence, the research focus needs to be directed to deriving models for predicting the stochastic behavior of human road users, while also avoiding collisions with them.

	This paper targets the autonomous driving problem in complex urban environments where road users are present. The autonomous vehicle needs to drive along a pre-defined  route, while also ensuring that collisions are avoided. 
	To that end, it is necessary that the vehicle is provided with information about the environment, such as where a road user will most likely be in the future, in order to be proactive and plan for a collision-free path. Therefore, the vehicle motion planning and control need to be handled in real-time, hence a low computational complexity is needed.
	
	While safety is the dominating factor when designing driving trajectories for autonomous driving, passengers' comfort also needs to be considered. The main contributing factors to ride discomfort are considered to be high levels of jerk and acceleration \cite{nilsson2014performance}. Several methods have been proposed where these factors are taken in consideration and optimized for driving scenarios on highways or country roads \cite{ziegler2014trajectory,nilsson2015longitudinal}. Similarly, the work in \cite{bianco2007optimal,bianco2013minimum} minimizes jerk profiles under fixed travel times set beforehand. 
	
	For urban driving, \cite{li2015practical} presents a framework for trajectory planning by decomposing the problem in spatial path planning and velocity profile planning. However, the planner does not consider reactive behaviors such as responding to dynamic objects in the environment. 
	Other frameworks \cite{benenson2006integrating,johnson2013optimal} use grid or graph-based methods to either plan paths around pedestrians or stop for them, while making an assumption that alternative paths exist, which may not always hold true in urban scenarios.  Planners using human comfort as optimization are also able to plan paths around obstacles \cite{villagra2012path,villagra2012smooth}, but avoid stopping as it is not always optimal. Similar to other planners, \cite{cofield2016reactive} uses a path-velocity decomposition to generate path and velocity profiles in parallel. While the planner is able to slow down, or come to a complete stop for pedestrians, it does not consider pedestrian movements in time, which might affect the planning capability.
	

	
	\begin{figure}[t!]
		\centering
		\mbox{\parbox{.48\textwidth}{
				\centering
				\includegraphics[width=\linewidth]{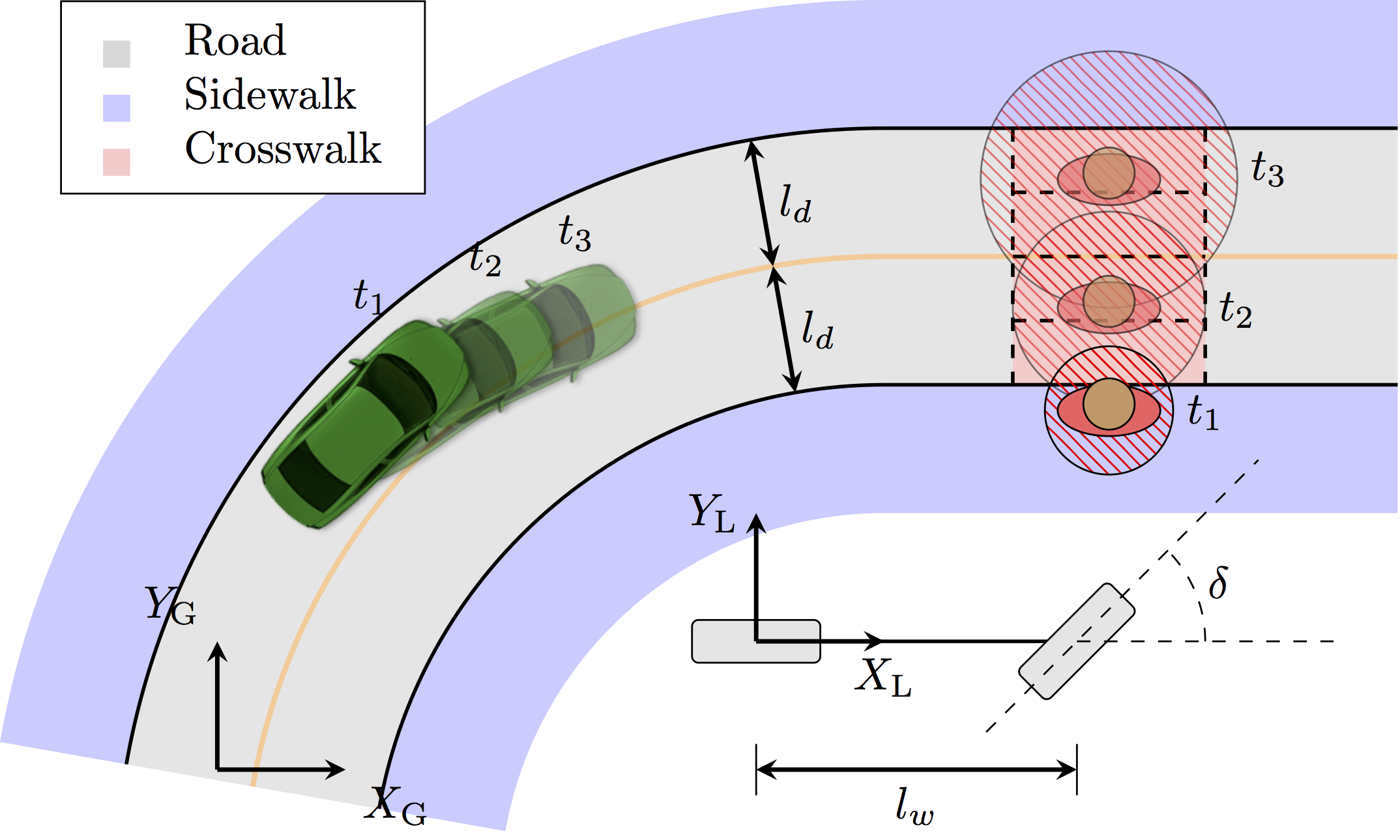}
		}}
		\caption{An illustration of a path planning problem. The vehicle needs to plan a trajectory in time subject to constraints such as road boundaries, but also moving pedestrians (red circles). The kinematic bicycle model is depicted in the lower part of the figure.
		}
		\label{fig:modelSketch}
		\vspace{-1.5em}
	\end{figure}

	In this paper we provide a general real-time framework to handle autonomous driving in urban scenarios with pedestrian crossing intersections. Unlike the common approach in literature, we do not decouple the longitudinal and lateral control of the vehicle. Instead, we solve the trajectory planning and vehicle control problem simultaneously, with the assumption that a higher level navigation layer provides a nominal driving route, e.g. center of the road lane, that the vehicle needs to track. Model Predictive Control~(MPC) is used to generate optimal trajectories, while considering other road users by predicting their motion in time. When considering other road users, other approaches in the literature often predict the future motion with primitive methods, e.g. constant velocity or acceleration models. In this paper we use an environment-aware predictor, developed in \cite{batkovic2018computationally}, to obtain more  accurate future predictions, and unify the collision avoidance problem simultaneously with the vehicle motion and control problem.

	We ensure collision avoidance by transforming the predicted motion and uncertainty of other road users into constraints within the MPC formulation. Doing so, allows us to plan a trajectory in time, while being proactive and avoiding collisions with moving road users or other static obstacles. Fig.~\ref{fig:modelSketch} shows an example of how predicted pedestrian positions can be used for collision avoidance. The red circles express regions at different planning times that the vehicle is not allowed to enter. It is important to note that we aim at addressing a nominal behavior which guarantees safety and comfort, but which also requires an emergency layer which would intervene in case something unexpected happens. The safety layer is the subject of ongoing research.
	
 	
	This paper is structured as follows. In Section~\ref{sec:problem} we introduce the problem formulation and our framework along with the used vehicle and pedestrian prediction models. The framework is evaluated, both in simulations and real experiments, for scenarios with simulated pedestrians moving near an intersection in Section~\ref{sec:simulations}. Finally, we draw conclusions and outline future research directions in Section~\ref{sec:conclusions}.

	\section{Problem Formulation and Framework} \label{sec:problem}
    An autonomous driving application needs above all to be safe, and in order to safely coexist with a higher level of riding comfort, the vehicle behavior needs to be proactive. 
    By predicting the 
    future evolution of the driving environment, one can incorporate cautiousness to minimize the risk of collisions with surrounding traffic participants (pedestrians, cyclists, cars). 
    In the following, we first introduce the system architecture, then we explain the vehicle and pedestrian models used in our decision problem. Finally, we present the reference path generation, cost function, and system constraints used to formulate our control objectives.
    
    
   \subsection{System Architecture} 
\begin{figure}[t]
    	\vspace{0.5em}
    	\centering
    	\mbox{\parbox{.48\textwidth}{
    			\centering
    			\includegraphics[width=\linewidth]{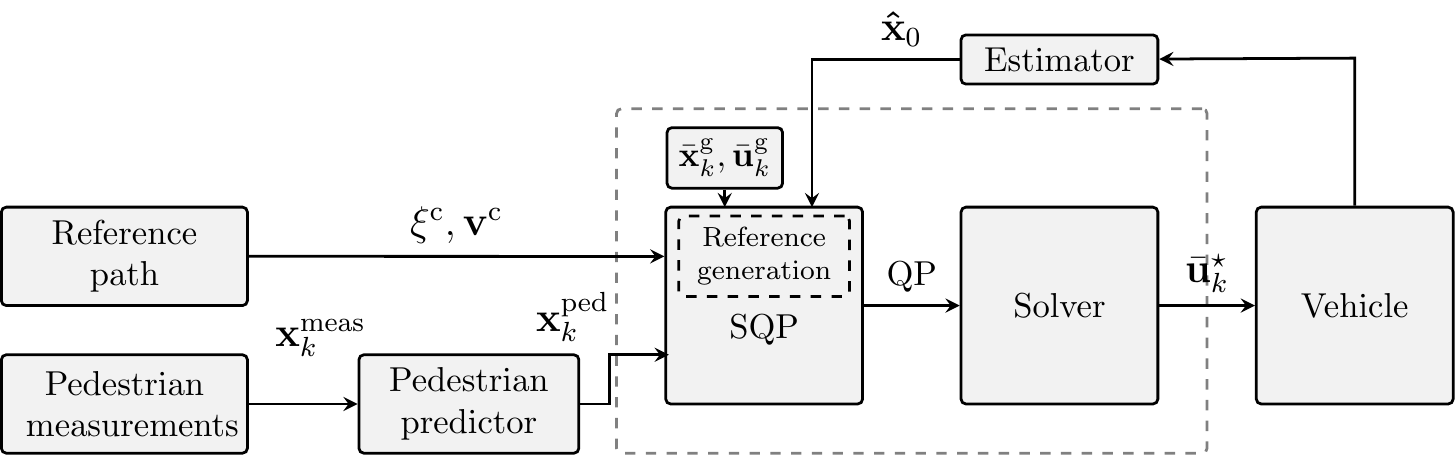}
    	}}
    	\caption{Representation of the closed loop architecture.
    	}
    	\label{fig:systemArchitecture}
    	\vspace{-1.5em}
    \end{figure}

    At each sampling instant, we assume that the vehicle receives estimated vehicle states $\hat\x_0$ from an estimator, previous guesses of state and control input trajectories $\bar\x_k^\mathrm{g}$ and $\bar\u_k^\mathrm{g}$ respectively, and a path reference $\mathbf{\xi}^\mathrm{c}$ and $\mathbf{v}^\mathrm{c}$ from a nominal path module. State measurements $\mathbf{x}_k^\mathrm{meas}$ of a detected pedestrian\footnote{Although our approach can be extended to other road users, for convenience of exposition we'll refer to pedestrians} are sent to a module that generates predictions $\x_k^\mathrm{ped}$ of the pedestrian positions in time. The planner uses the estimated states, the state and control guesses, the path references, and the pedestrian predictions to control the vehicle along the path, while avoiding collisions. This paper focuses on the planner represented as the largest dashed rectangle in Fig.~\ref{fig:systemArchitecture}.
    

    \subsection{Vehicle Model}
    For urban autonomous driving, lateral, longitudinal, and yaw dynamics are quite fast \cite{rajamani2011vehicle}. 
    Therefore, a kinematic bicycle model was chosen for its simplicity and comparable accuracy with a dynamic one \cite{kong2015kinematic} in the scenarios considered in this paper. A vehicle model sketch is shown in Fig.~\ref{fig:modelSketch}. 
    
    We model the vehicle motion with states $\mathbf{x}\in\mathbb{R}^6$, and controls $\mathbf{u}\in\mathbb{R}^2$, defined as
   \begin{align*}
   \x := \matr{cccccc}{x & y& v& \theta&\delta & \omega}^\top, && \u := \matr{c c}{a & \delta_\mathrm{sp}}^\top,
   \end{align*}
    where we denote the position coordinates of the rear wheel in the absolute reference frame G as $x, y$, the velocity in direction $\theta$ as $v$, and the steering angle and steering angle rate as $\delta$ and $\omega$, respectively. The acceleration and steering angle setpoint are denoted respectively by $a$ and $\delta_\mathrm{sp}$, where the setpoint refers to desired steering  angle at the wheel base.
    
    With the chosen state representation, the dynamics are described by the bicycle equations as
    \begin{equation}
    \label{eq:kinematic}
    \dot{\x} = \matr{c}{\dot{x} \\ \dot{y} \\ \dot{v} \\ \dot{\theta} \\ \dot{\delta} \\ \dot{\omega}}
    = \matr{c}{v\cos\theta \\ v\sin\theta \\ a \\ v/l_w\tan\delta \\ \omega \\ w_0^2(\delta_\mathrm{sp}-\delta)-2\zeta\omega}=: f_\mathrm{c}(\x,\u),
    \end{equation} 
    where the only model parameters are: the wheel base $l_w$ and the parameters $w_0$ and $\zeta$, introduced to describe a second order model of the steering wheel actuator. 
    
    For simplicity, we chose to omit the longitudinal jerk in the model. It could, however, easily be implemented without any greater complication or performance impact.

    \subsection{Pedestrian Model}
    \label{sec:pedestrianModel}
    
    The performance of a planner is predominantly limited by the accuracy of the information it is provided with: generating plans with highly inaccurate information about the surrounding environment will result in suboptimal and possibly unsafe or conservative plans.
    Since our planner solves an optimal control problem over a prediction horizon, a description of the environment, e.g. pedestrian positions and road boundaries, is necessary throughout the horizon.
    
    We chose to use the flexible and fairly accurate method presented in \cite{batkovic2018computationally} to generate pedestrian predictions needed for our planner. A simple representation of the road geometry is used in order to assign possible paths (hereafter referred to as references), e.g. sidewalks and crosswalks, where a pedestrian might walk. The pedestrian motion is modeled using a unicycle model, which is assumed to follow the reference, thanks to a closed-loop regulator. 
    Process noise is added to the model, and state uncertainty is predicted by propagating its covariance. 
    Using the closed-loop system model, pedestrian trajectories and state covariances can be predicted along the road geometry. 
    
    Moreover, if a pedestrian is walking on a road that has a bifurcation, e.g. the road splits into multiple walking paths, the prediction method takes into account all possible directions and generates predictions along these paths. As the number of paths or obstacles increase, the computational complexity may be contained through parallelization.
    
    

   \subsection{Problem Statement}
   The objective of the vehicle is to safely follow any given path as close as possible, while driving comfortably and satisfying constraints arising from the physical limitations, but also other road users, e.g pedestrians, cyclists and other vehicles.
   
   This problem can be formulated as a finite horizon, constrained  optimal control problem. Since the vehicle model and constraints are nonlinear, we are faced with the following Nonlinear Model Predictive Control~(NMPC) problem
    \begin{subequations}
    	\label{eq:nmpc}
		\begin{align}
		\min_{\bar\x,\bar\u} & \sum_{k=0}^{N-1}
		\matr{c}{\bar\x_k - \r_k^\x \\ \bar\u_k - \r_k^\u}^\top W_{k} \matr{c}{\bar\x_k - \r_k^\x \\ \bar\u_k - \r_k^\u} \\
		&\qquad + \matr{c}{\bar\x_N - \r_N^\x}^\top W_N \matr{c}{\bar\x_N - \r_N^\x}\nonumber\\
		&\text{s.t.}\ \ \, \bar\x_0 = \hat{\x}_0, \label{eq:nmpcState}\\
		&\qquad{}\bar\x_{k+1} = f(\bar\x_{k},\bar\u_{k}),\label{eq:nmpcDynamics}\\
		&\qquad{}h(\bar\x_k,\bar\u_k) \leq{} 0, \label{eq:npmcInequality}
		\end{align}
	\end{subequations}
   where $N$ is the prediction horizon, $W_k$ is the stage cost matrix, $\bar\x_k$ and $\bar\u_k$ are the state and control input, $\r_k^\x$ and $\r_k^\u$ are the state and control input reference, constraint \eqref{eq:nmpcState} enforces that the prediction starts at the current state estimate $\hat\x_0$, constraint \eqref{eq:nmpcDynamics} enforces the system dynamics, and \eqref{eq:npmcInequality} enforces constraints such as, e.g. actuator limits and obstacle avoidance. The system dynamics in discrete time are obtained through numerical integration of \eqref{eq:kinematic}. Note that proactivity naturally follows in the MPC framework through consideration of predicted constraints over the horizon.
   
   With the algorithms described in \cite{gros2016linear,diehl2005real} the Nonlinear Program~(NLP) \eqref{eq:nmpc} can be rewritten using a Sequential Quadratic Programming~(SQP) approach, which sequentially approximates the NLP with a Quadratic Program~(QP) \eqref{eq:qp}. Here, we only give a brief summary of the methodology. For any further details we refer the reader to \cite{gros2016linear,diehl2005real}.
   Following the SQP approach we linearize the system model with a previous guess $(\bar\x_k^\mathrm{g},\bar\u_k^\mathrm{g})$ and use any numerical discretization scheme to obtain the sensitivities
   \begin{equation*}
   A_k = \frac{\partial{}f}{\partial{}\x}\bigg\rvert_{\bar\x_k^\mathrm{g},\bar\u_k^\mathrm{g}}, B_k = \frac{\partial{}f}{\partial{}\u}\bigg\rvert_{\bar\x_k^\mathrm{g},\bar\u_k^\mathrm{g}}, b_k = f(\bar\x_k^\mathrm{g},\bar\u_k^\mathrm{g}) - \bar\x_{k+1}^\mathrm{g},
   \end{equation*}
   for every prediction time step $k$. The quantities $C_k$, $D_k$, $\mathbf{d}_k^\mathrm{l}$, $\mathbf{d}_k^\mathrm{u}$ $C_N$, $\mathbf{d}_N^\mathrm{l}$, and $\mathbf{d}_N^\mathrm{u}$ can be computed in a similar manner from constraint \eqref{eq:npmcInequality}. With the sensitivities, the problem is then formulated into a QP and solved, where the solution is used to update the previous guess. This process is either iterated until convergence, or only done once, thus following the real time iteration~(RTI) scheme \cite{diehl2005real}. At each iteration, the following QP is solved

    \begin{subequations}
   	\label{eq:qp}
   	\begin{align}
   	\min_{\bar\x,\bar\u} & \sum_{k=0}^{N-1}
   	\matr{c}{\bar\x_k - \r_k^\x \\ \bar\u_k - \r_k^\u}^\top \matr{cc}{Q_k & S_k \\ S_k^\top & R_k} \matr{c}{\bar\x_k - \r_k^\x \\ \bar\u_k - \r_k^\u} \\
   	&\qquad + \matr{c}{\bar\x_N - \r_N^\x}^\top Q_N \matr{c}{\bar\x_N - \r_N^\x}\nonumber\\
	&\text{s.t.}\ \ \, {}\bar\x_0 = \hat{\x}_0, \\
    &\qquad\bar\x_{k+1} = A_k \bar\x_k + B_k \bar\u_k + b_k,\label{eq:qpDynamics}\\
	&\qquad{}\mathbf{d}^\mathrm{l}_k \leq{} C_k\bar\x_k + D_k\bar\u_k \leq{} \mathbf{d}^\mathrm{u}_k \label{eq:qpStage}\\
	&\qquad{}\mathbf{d}^\mathrm{l}_N \leq{} C_N\bar\x_N  \leq{} \mathbf{d}^\mathrm{u}_N. \label{eq:qpFinal}
    \end{align}
   \end{subequations}
   The tuning parameters here are the stage cost matrices $Q_k$, $R_k$ and $S_k$, and the prediction horizon $N$. Note that all matrices in this formulation are in general time-varying.
   
   At every sampling instant, Problem \eqref{eq:qp} is solved, and only the first control input $\u_0$ is applied to the system. However, the solutions $\bar\x$ and $\bar\u$ are used to update the guesses $(\bar\x_k^\mathrm{g},\bar\u_k^\mathrm{g})=(\bar\x_{k+1},\bar\u_{k+1})$, where $k$ spans the prediction horizon, following the  shifting approach in \cite{gros2016linear,diehl2005real}. 


   \begin{figure}[t!]
   	\vspace{.4em}
   	\centering
   	\mbox{\parbox{.48\textwidth}{
   			\centering
   			\includegraphics[width=\linewidth]{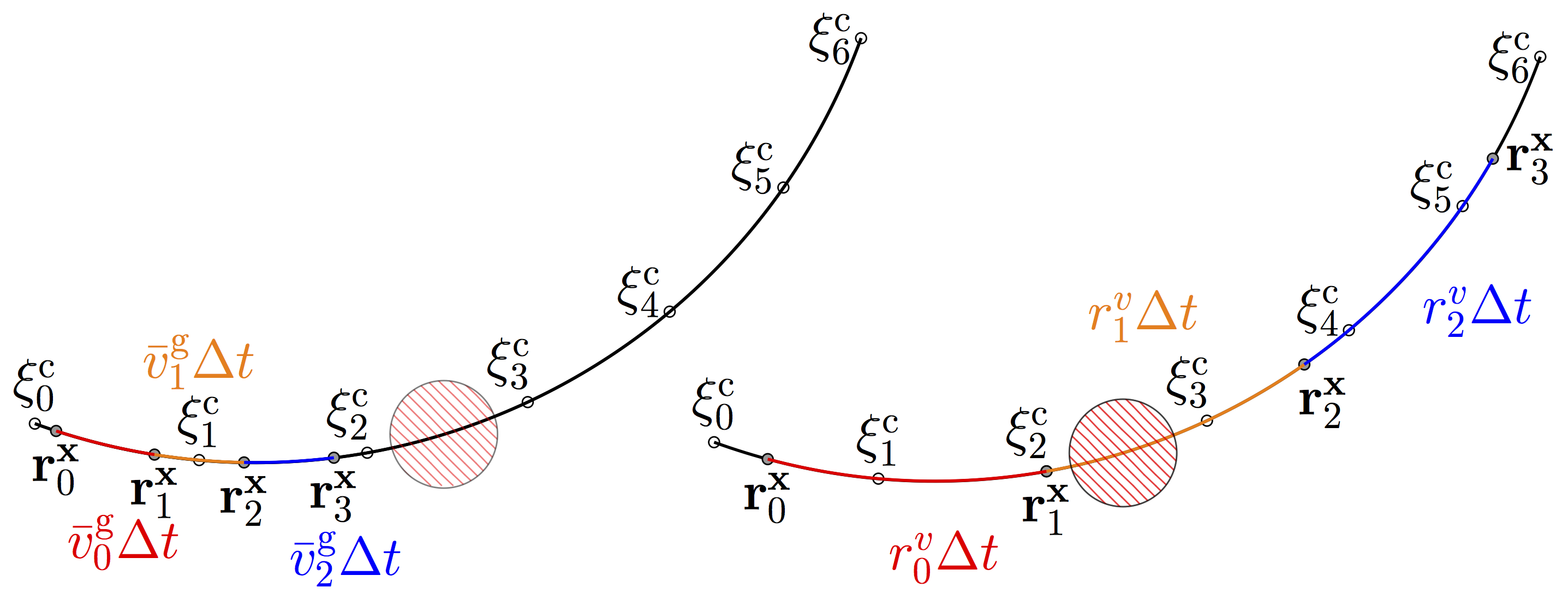}
   	}}
   	\caption{Illustration of generated reference points $\r_k^\x$. The left and right curves show the reference points obtained by integration with either the previous velocity profile $\bar{v}_k^\mathrm{g}$ or  the velocity reference $r_k^v$. Using the velocity reference  $r_k^v$, the generated reference points $\r_k^\x$ do not account for the obstacle (red circle).
   	}
   	\label{fig:sigmaRef}
   	\vspace{-1.5em}
   \end{figure}
   
   \subsection{Reference Generation}
   \label{sec:reference}
		The generation of the trajectory reference deserves a specific discussion, since it is a crucial component of the control scheme and, as such, has a strong effect on the closed-loop behavior. 
		In standard MPC implementations, the reference is updated from one time step to the next by shifting: $(\r_k^\x,\r_k^\u)=(\r_{k+1}^\x,\r_{k+1}^\u)$. In our setup instead, we recompute the reference at every time instant in order to avoid aggressive controller behaviors when the vehicle is forced to slow down due to the presence of an obstacle.
		 
		 In particular, rather than relying on a reference trajectory, we choose a reference path and a reference velocity along it. The advantage of such a choice is that, if the vehicle is forced to slow down by the presence of an obstacle, the controller will not try to recover the lost time and instead it will simply try to bring the vehicle back at the desired velocity. In our research, we adopt an approach which is very similar to the one proposed in~\cite{Faulwasser2009} to address these issues. Differently from~\cite{Faulwasser2009}, we eliminate the curvilinear path parameter $\sigma$ from the problem formulation and we relate $\dot \sigma$ to the vehicle's velocity directly.
		
		We assume that a reference velocity $v^\mathrm{c}(\sigma)$ and a curve $\xi(\sigma)=(x^\mathrm{c}(\sigma),y^\mathrm{c}(\sigma),\theta^\mathrm{c}(\sigma),\delta^\mathrm{c}(\sigma),\omega^\mathrm{c}(\sigma),a^\mathrm{c}(\sigma),\delta_\mathrm{sp}^\mathrm{c}(\sigma))$ parametrized in the curvilinear coordinate $\sigma$ are available. Using the current position $(x_0,y_0)$, we define the initial reference coordinate $\sigma_0$ by projecting $(x_0,y_0)$ onto the curve $\xi(\sigma)$ such that $\sigma_0 = \mathrm{arg}\min_\sigma \| (x_0,y_0) - (x^\mathrm{c}(\sigma),y^\mathrm{c}(\sigma)) \|_2$. Then, we define reference points at time $k\Delta{t}$, where $\Delta{t}$ is the discretization time, by means of $\sigma_k=\sigma(k\Delta{t})$. In order to obtain a meaningful reference in the presence of obstacles, 
		we propose to use
		\begin{equation}
		 \sigma_k=\int_0^t(1-\kappa(\tau)e_y(\tau))^{-1}v(\tau)\cos(e_\psi(\tau))d\tau,
		 \end{equation}
		 where $\kappa$, $e_y$, and $e_\psi$ is the curvature, lateral error and orientation error w.r.t to the path. This guarantees that the predicted reference accommodates for the presence of obstacles \cite{verschueren2016time}. Using the reference velocity in the computation of $\sigma(t)$, instead, would ignore the presence of obstacles and create the potential for unstable behavior, especially in the presence of sharp turns. 
		In the context of the RTI scheme, $\sigma(t)$ is obtained by integration of the velocity computed at the previous time instant. 
		
		
		In practice, the curves $\xi(\sigma)$ and $v(\sigma)$ might not be available and the reference generation layer might have access only to a set of data points $\xi^\mathrm{c}_p$ and $v_p^\mathrm{c}$. Then, a curve can be obtained by interpolation. In this paper, we use a simple piecewise linear interpolation and the formula 
		$\sigma_{k+1}=\sigma_k + \bar{v}_k^\mathrm{g}\cos(e_{\psi_k})\Delta{t}$ for  approximate integration, where $\bar{v}_k^\mathrm{g}$ is the velocity component of $\bar\x^\mathrm{g}_k$. 
		By selecting $\sigma_k$ according to $\bar{v}_k^\mathrm{g}$ rather than $r_k^{v}$, the algorithm is able to account for the presence of obstacles, as explained in Fig.~\ref{fig:sigmaRef}. Note that in the second case, the orientation reference will refer to a point on the path far from the previously predicted vehicle position and, therefore, will result in an incorrect orientation reference and unwanted driving behavior.
		
		Moreover, we assume that the only available data concerns positions and velocities, such that we need to define suitable references for the other states and controls. The position references are obtained as $(r_k^x,r_k^y)=(x^c(\sigma_k),y^c(\sigma_k))$, while the heading reference $r^\theta_k$ is computed by
		\begin{equation}
		r_k^\theta = \tan^{-1}\Big(\frac{r_{k+1}^y-r_{k}^y}{r_{k+1}^x-r^x_k}\Big),
		\end{equation} 
		and the steering angle and steering angle rate references are set as the initial angle $\delta_0$ and zero, respectively, across all prediction times $k$. The reference for the control input is set as zero for the acceleration, and $\delta_0$ for the steering setpoint.
   \begin{figure}[t!]
	\vspace{0.5em}
   	\centering
   	\mbox{\parbox{.48\textwidth}{
   			\centering
   			\includegraphics[width=1\linewidth]{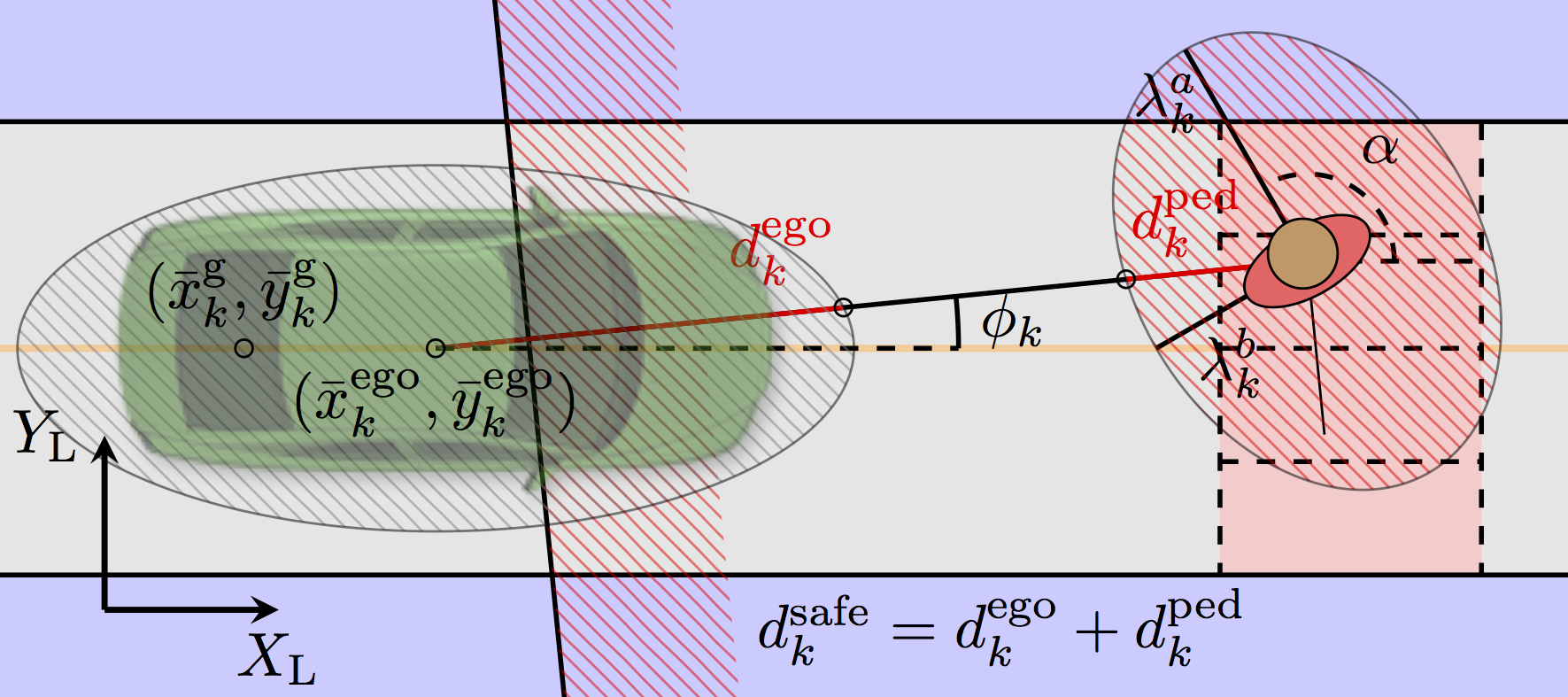}
   	}}
   	\caption{Illustration of the regions occupied by the vehicle and pedestrian. A linear constraint is expressed by the black line, distancing itself with distance $d_k^\mathrm{safe}$ from the pedestrian.
   	}
   	\label{fig:linConstr}
   	\vspace{-1.5em}
   \end{figure}
   
   \subsection{Cost Function}
		Often, matrices $Q_k$, $R_k$, and $S_k$ are chosen to have the diagonal form
		\begin{equation}
		\begin{gathered}
		Q_k = \mathrm{diag}(\bar{q}_1, ..., \bar{q}_m),\\
		R_k = \mathrm{diag}(\bar{r}_1, ..., \bar{r}_n),\quad S=0,
		\end{gathered}
		\end{equation}
		where $m$ and $n$ denote the state and control input dimensions.
		
		As mentioned in Section~\ref{sec:reference}, in the presence of obstacles, trying to force the system to catch up with a predefined evolution of the curvilinear coordinate $\sigma$ can yield an undesirably aggressive behavior. Therefore, we propose to remove the term penalizing this deviation from the cost. We do so by only penalizing the lateral deviation from the path $\xi^\mathrm{c}$ through the term
		\begin{align*}
		\left (\matr{c}{0\\1}^\top\matr{cc}{\cos(-r^\theta_k)& -\sin(-r^\theta_k)\\\sin(-r^\theta_k) & \cos(-r^\theta_k)}\matr{c}{\bar{x}_k-r_k^x\\ \bar{y}_k-r_k^y}\right )^2.
		\end{align*}
		Then, matrix $Q_k$ becomes
		\begin{equation}
		Q_k=\mathrm{blockdiag}(\bar{q}_1T(-r^\theta_k),\ \bar{q}_2,\ \bar{q}_3,\ \bar{q}_4,\ \bar{q}_5),
		\end{equation}
		with
		\begin{equation}
		T(\theta)= \matr{cc}{\sin^2\theta& \cos\theta\sin\theta\\
			\cos\theta\sin\theta & \cos^2\theta }.
		\end{equation}
		With this cost formulation we avoid undesirably aggressive behaviors and we still ensure that the cost matrices are positive semi-definite.

   \subsection{System Constraints}
   The states and control inputs are subject to a set of constraints formulated in \eqref{eq:qpStage}-\eqref{eq:qpFinal}, which concern mainly safety, comfort, and physical limitations of the vehicle. These consist of the road boundary limits, but also of occupied areas of predicted pedestrians' positions. 
   
   \subsubsection{Road Boundary Constraints} In order to decide on the lateral bounds of the road, we use the previous guesses $\bar\x^\mathrm{g}$ to project the positions $(\bar{x}^\mathrm{g}_k,\bar{y}^\mathrm{g}_k)$ onto the path. Knowing the projection point $(\hat{x}^\mathrm{g}_k,\hat{y}^\mathrm{g}_k)$, and orientation $\hat{\theta}_k^\mathrm{g}$,  it is possible to express a lateral bound $l_d$ on the path as
   \begin{equation}
    \Delta^\mathrm{road}-l_d \leq{} \bar{x}_k\sin(-\hat{\theta}_k^\mathrm{g})+\bar{y}_k\cos(-\hat{\theta}_k^\mathrm{g}) \leq{} \Delta^\mathrm{road}+l_d, \label{eq:roadConstr}
   \end{equation}
   where $\Delta^\mathrm{road}=\hat{x}_k^\mathrm{g}\sin(-\hat{\theta}_k^\mathrm{g})+\hat{y}_k^\mathrm{g}\cos(-\hat{\theta}_k^\mathrm{g})$.

    \subsubsection{Pedestrian Constraints}
    Information about future pedestrian states is provided by a prediction layer outside of the MPC framework. The predictions consist of positions $(x_k^\mathrm{ped},y_k^\mathrm{ped})$, the major and minor axes $\lambda_k^a$ and $\lambda_k^b$ corresponding to the covariance ellipse of a Gaussian probability density function, and the orientation $\alpha$ in a global frame. With this information, occupied pedestrian regions are defined in the time domain. Similarly, we define a safety ellipse around the vehicle that describes the space it occupies. An illustration is provided in Fig.~\ref{fig:linConstr}.
    
    The covariance ellipses  are linearized using the previous vehicle positions $(\bar{x}_k^\mathrm{g},\bar{y}_k^\mathrm{g})$ and heading $\bar\theta_k^\mathrm{g}$ to find the vehicle center $(\bar{x}_k^\mathrm{ego},\bar{y}_k^\mathrm{ego})$. The  predicted pedestrian positions $(x_k^\mathrm{ped},y_k^\mathrm{ped})$ and the vehicle center are used to compute the safety distances $d_k^\mathrm{ego}$ and $d_k^\mathrm{ped}$ for the vehicle and  pedestrian using the relative orientation $\phi_k$. Considering the two safety distances, we express a linear constraint around the rear wheel of the vehicle as

    \begin{equation}
    \Delta^\mathrm{ped}\mathbf{l}_b \leq{} \Delta^\mathrm{ped}\matr{c}{\bar{x}_k\\ \bar{y}_k},
    \end{equation}
    where $\Delta^\mathrm{ped}=-\matr{cc}{\cos\phi_k&\sin\phi_k}$ and
    \begin{equation}
	\mathbf{l}_b = \matr{c}{x_k^\mathrm{ped} - d_k^\mathrm{safe}\cos\phi_k -(\bar{x}_k^\mathrm{ego}-\bar{x}_k^\mathrm{g})\\
	y_k^\mathrm{ped} - d_k^\mathrm{safe}\sin\phi_k -(\bar{y}_k^\mathrm{ego}-\bar{y}_k^\mathrm{g})}.
    \end{equation} 
    The  resulting linearized constraint, together with a physical illustration of the parameters, is illustrated in Fig.~\ref{fig:linConstr}.

	\section{Simulations \& Experimental Validation} \label{sec:simulations}
	In this section we present results from a traffic scenario simulation with a crossing pedestrian, and the ego vehicle. To model the pedestrian, we used a trajectory from the data set in \cite{batkovic2018computationally}. To further verify the framework, we also provide experimental results for a traffic intersection with a simulated pedestrian.
	
	
	\subsection{Simulation Setup}
	The controller was first tested in a simulated intersection scenario with a real pedestrian trajectory taken from \cite{batkovic2018computationally}. In this scenario, the ego vehicle is traveling along a straight road towards an intersection which a pedestrian is  approaching, see Fig. \ref{fig:scenarioPedestrian2}.
	
	
	We generated sensitivites $A_k$ and $B_k$ and $b_k$ with a fourth order Runge-Kutta method, while ensuring that 5 integrator steps was sufficiently accurate, with a  discretization time of  $\Delta{}t = 0.05$s, and the following parameter values
	\begin{equation}
	l_w = 2.984  \mathrm{m}, \quad w_0= 20\mathrm{s}^{-1},  \quad \zeta = 0.9\mathrm{s}^{-1}.\label{eq:paramValues}
	\end{equation}
	The estimated  state $\hat\x_0$ was obtained by integration, using an integrator with error control to guarantee the accuracy of the  solution. The  vehicle controller used
	\begin{equation}
	\begin{gathered}
	Q_k=\mathrm{blockdiag}(2T(-r^\theta_k),\ 0.1,\ 10,\ 0.1,\ 10),\\
	R_k = \mathrm{blockdiag}(2, 1),\ \quad S_k = 0,\ N = 100,\label{eq:tuningValues}
	\end{gathered}
	\end{equation}
	and the state and control inputs were subject to the following constraints
	\begin{equation}
	\begin{gathered}
	-1\leq{}\bar{v}_k\leq{}20,\ |\bar\delta_k|\leq{}0.4942,\ |\bar\omega_k|\leq{}0.1765,\\
	-2\leq{}\bar{a}_k\leq{}1,\ |\bar\delta_{k,\mathrm{sp}}|\leq{}0.4942,\label{eq:constraints}
	\end{gathered}
	\end{equation}
	where all values are given in standard SI-units.
	
	Finally, the lateral bound on the road boundary constraint \eqref{eq:roadConstr} was set to be $l_d=1$m. In order to ensure feasibility of optimization problem \eqref{eq:qp}, we relax constraint \eqref{eq:roadConstr} with an L1-penalty. This ensures that if there exists a feasible solution for problem \eqref{eq:qp}, then the relaxed problem yields the same solution \cite[Prop. 6]{hultoptimal}.

	\subsection{Experimental Setup}
\begin{figure}[t!]
		\vspace{0.5em}
		\centering
		\mbox{\parbox{.48\textwidth}{
				\centering
				\includegraphics[width=0.85\linewidth]{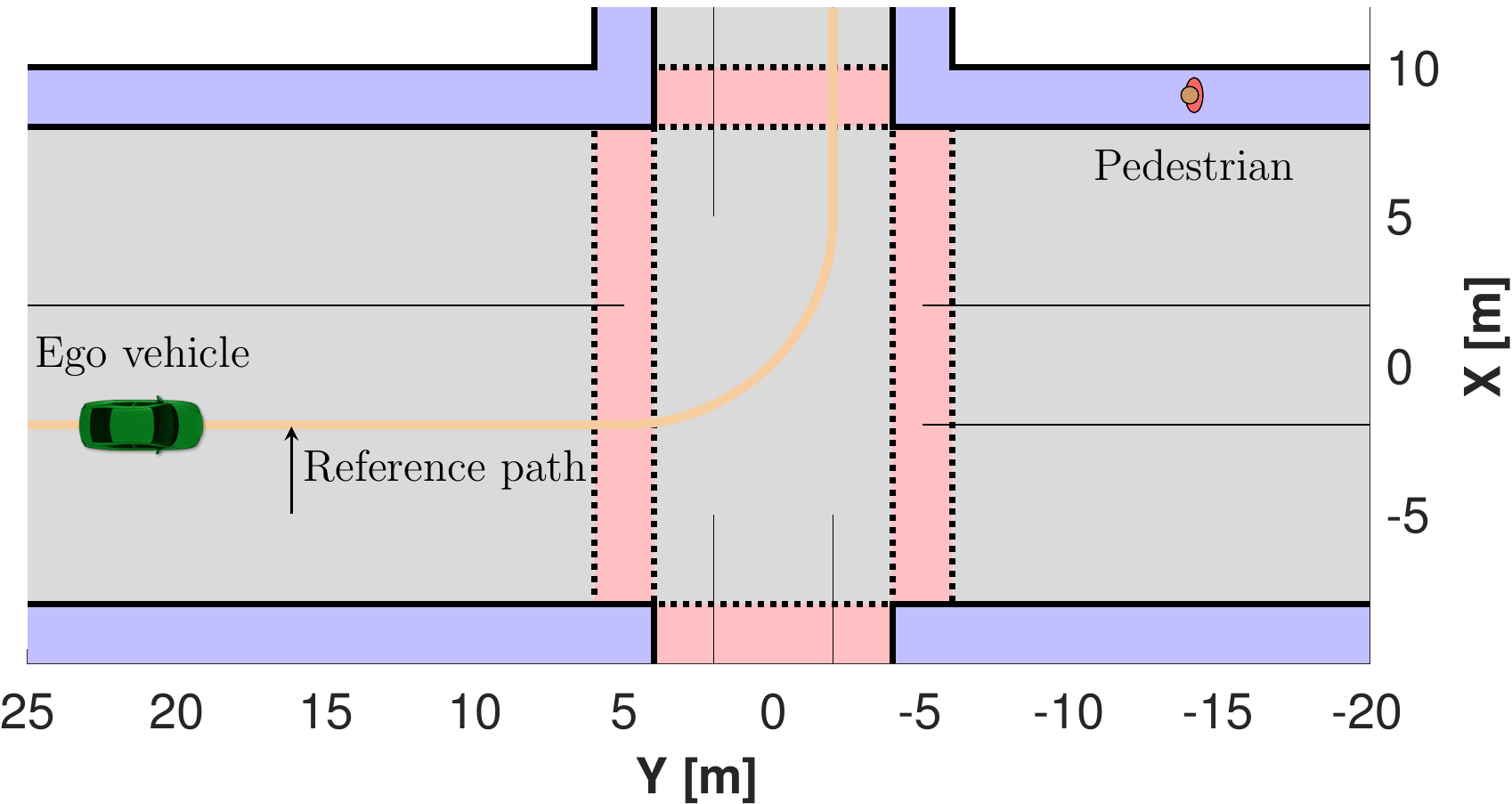}
		}}
		\caption{The vehicle needs to drive along its given path, and is faced with a pedestrian moving towards the intersection, not knowing if the pedestrian will cross or not.
		}
		\label{fig:scenario}
		\vspace{-1.5em}
	\end{figure}

	The closed loop controller was experimentally validated at the Astazero\footnote{See http://www.astazero.com/ for more information} test-track outside Gothenburg, Sweden. The experiment  consisted of a traffic scenario including our ego vehicle, and a simulated walking pedestrian. A simulated pedestrian was used due to safety reasons. The vehicle needed to make a 90-degree left turn in the intersection, while a simulated pedestrian was moving close-by and could possibly cross. Fig.~\ref{fig:scenario} shows the considered scenario, where the vehicle needs to drive along the orange line that represents the center of the driving lane, with an approaching pedestrian from the right.

	\subsubsection{Test  Vehicle} The vehicle used  in the experiment was a Volvo XC90 T6 petrol-turbo SUV with the real-time open source software OpenDLV\footnote{See https://opendlv.org/ for more information.} that interfaced sensors and actuators to an  external computer for control. A Laptop computer (i7 2.8GHz, 16GB RAM) running Ubuntu 16.04 as a Virtual Machine was used to receive sensor data, compute the control signals, and send actuation requests to the vehicle with OpenDLV through an ethernet connection. The vehicle was equipped with a Real Time Kinematic (RTK) GPS receiver for high accuracy positioning, and the estimated state $\hat\x{}_0$ was obtained with an Extended Kalman Filter (EKF).
		
	
	
	\subsubsection{Experimental Details} The interface between OpenDLV and the vehicle exhibited a delay of roughly $300$ms when sending signals to the steering actuator, but also reading from it. Therefore the state space vector in \eqref{eq:kinematic} was augmented with additional time-delayed states for the steering  angle to model the input delay. Since we estimated the steering wheel actuator dynamics, dead reckoning was used for estimating the steering wheel angle and steering wheel angle rate. Furthermore, due to safety features in the vehicle, the actuator limited the steering wheel angle in a speed-dependent way while in autonomous driving mode. To model this, we implemented the linear constraint
	\begin{equation}
	a_1\bar{v}_k + b_1\bar\delta_k \leq{} c_1, \quad a_2\bar{v}_k + b_2\bar\delta_k \geq{}c_2.
	\end{equation}
	
	The sensitivities were generated in the same way as for the simulation, and used same parameter values as in \eqref{eq:paramValues} and tuning parameters as  in \eqref{eq:tuningValues}.
	
	\subsection{Results}
	We implemented the proposed framework in MATLAB first for the simulations, and then auto generated the code into C++, where it was interfaced with OpenDLV for experimental testing.
	Both implementations used the open source software ACADO \cite{quirynen2017lifted} for auto generation of the sensitivities, and the solver from HPMPC \cite{frison2014high} to solve the QP problem with the RTI scheme.
	
	
	\subsubsection{Simulation}
\begin{figure}[t!]
	\vspace{0.5em}
	\centering
	\mbox{\parbox{.48\textwidth}{
			\centering
			\includegraphics[width=0.85\linewidth]{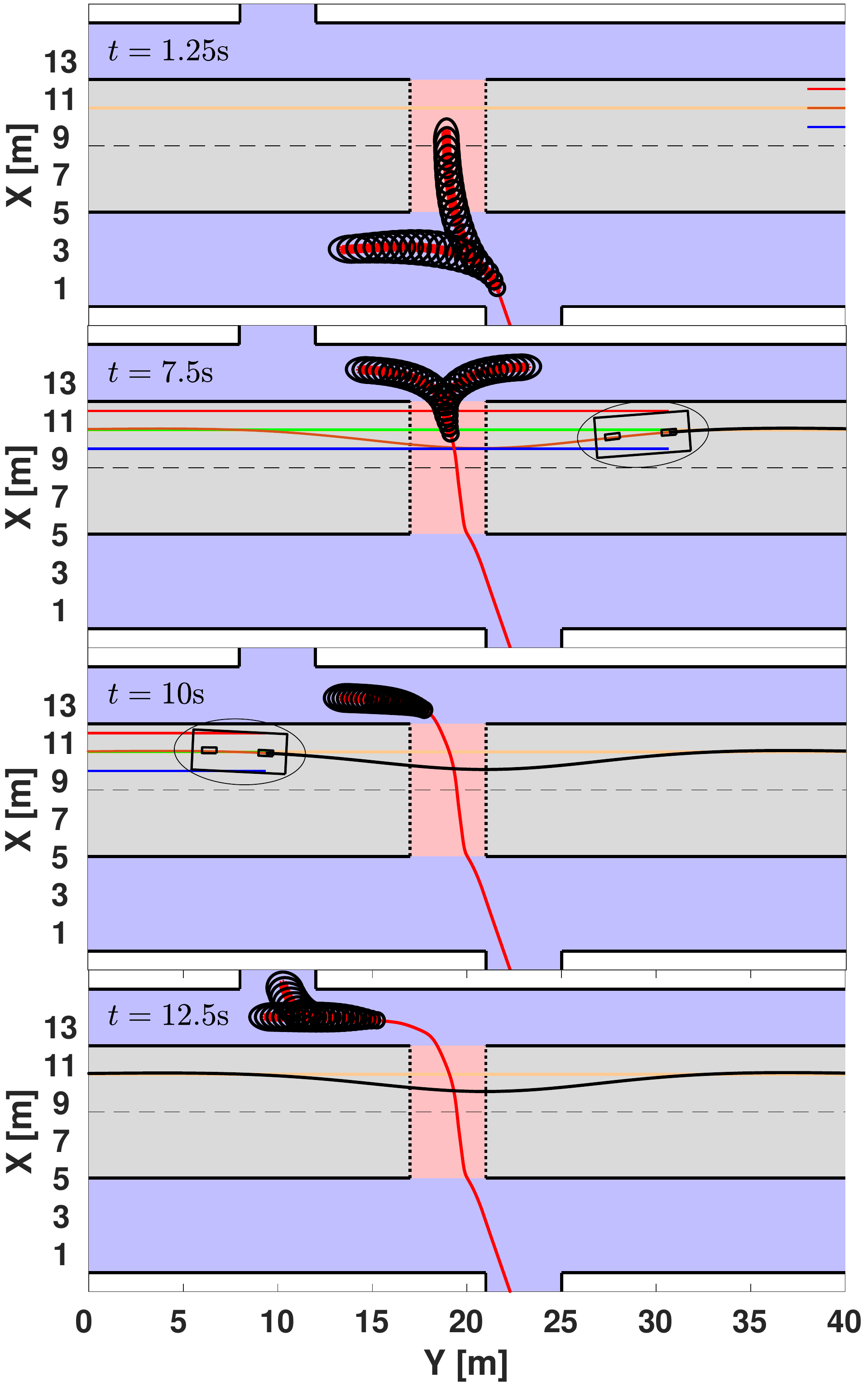}
	}}
	\caption{Scenario with real pedestrian measurements. The blue and red lines are the road boundary constraints from \eqref{eq:roadConstr}, while the green line and brown lines are the projected reference $(r_k^x,r_k^y)$ and open-loop solution $\bar\x$. The predicted pedestrian states are depicted as red points, and the uncertainties as black ellipses. The trailing black and red lines show the traveled path of the vehicle and pedestrian.
	}
	\label{fig:scenarioPedestrian2}
	\vspace{0em}
\end{figure}

\begin{figure}[t!]
	\centering
	\mbox{\parbox{.48\textwidth}{
			\centering
			\includegraphics[width=0.85\linewidth]{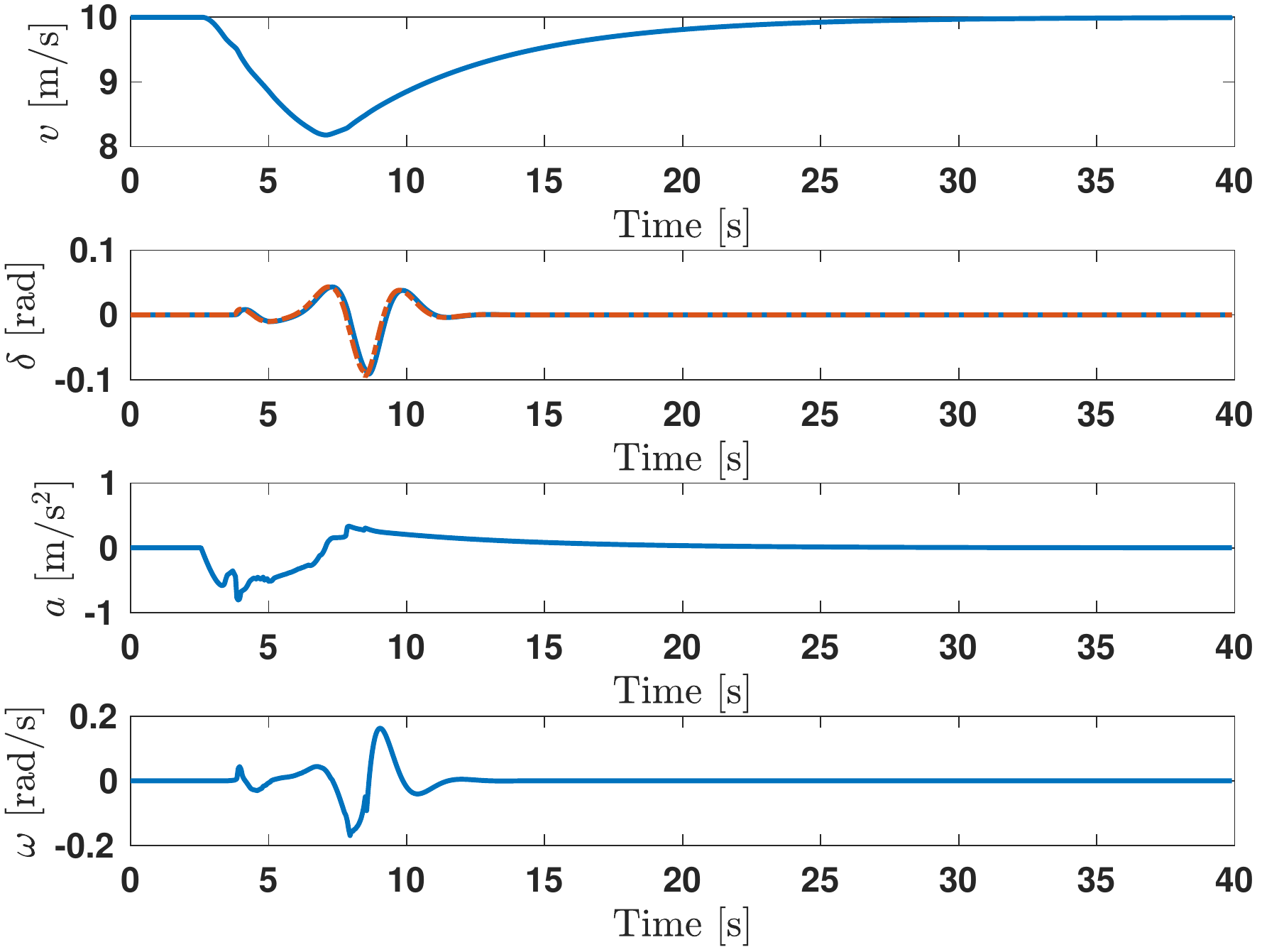}
	}}
	\caption{Closed loop states from Fig. \ref{fig:scenarioPedestrian2}. The first plot shows the velocity profile and the second plot shows the steering angle (blue line), and steering angle setpoint (dashed brown line). The last two plots show the acceleration and steering angle rate. 
	}
	\label{fig:scenarioPedestrianStates2}
	\vspace{-1.5em}
\end{figure}

	Fig.~\ref{fig:scenarioPedestrian2} illustrates the open-loop solutions across different time instants. The MPC  controller is generating control inputs to minimize the deviation from the reference path and reference velocity $r_k^v=10\mathrm{m}/\mathrm{s}$. For time $t=1.25$s the open-loop solution is not affected by the predicted pedestrian positions, and the controller is  generating inputs just to stay on the path. At time $t=7.5$s the vehicle has adapted its speed and managed to plan a path around the pedestrian instead of coming to a full stop. For times $t=10$s and $t=12.5$s, we can see how the vehicle finally passes the pedestrian, and the final closed-loop trajectory. The corresponding closed-loop states and control inputs are presented in Fig.~\ref{fig:scenarioPedestrianStates2}.

	The aggressive behavior on the steering rate is due to the deviation of the actual pedestrian position from the predicted one. When the vehicle is close to the pedestrian, the control authority is reduced and such uncertainties result in large control actions. Future research will investigate approaches to mitigate these undesirable behaviors.

	Lastly, the runtime of the framework in MATLAB resulted in an average solution time of $22.4$ms with a standard deviation of $2.3$ms, and longest timing of $34.1$ms. 
	
	\subsubsection{Experiment}
\begin{figure}[t!]
		\vspace{0.5em}
		\centering
		\mbox{\parbox{.48\textwidth}{
				\centering
				\includegraphics[width=0.85\linewidth]{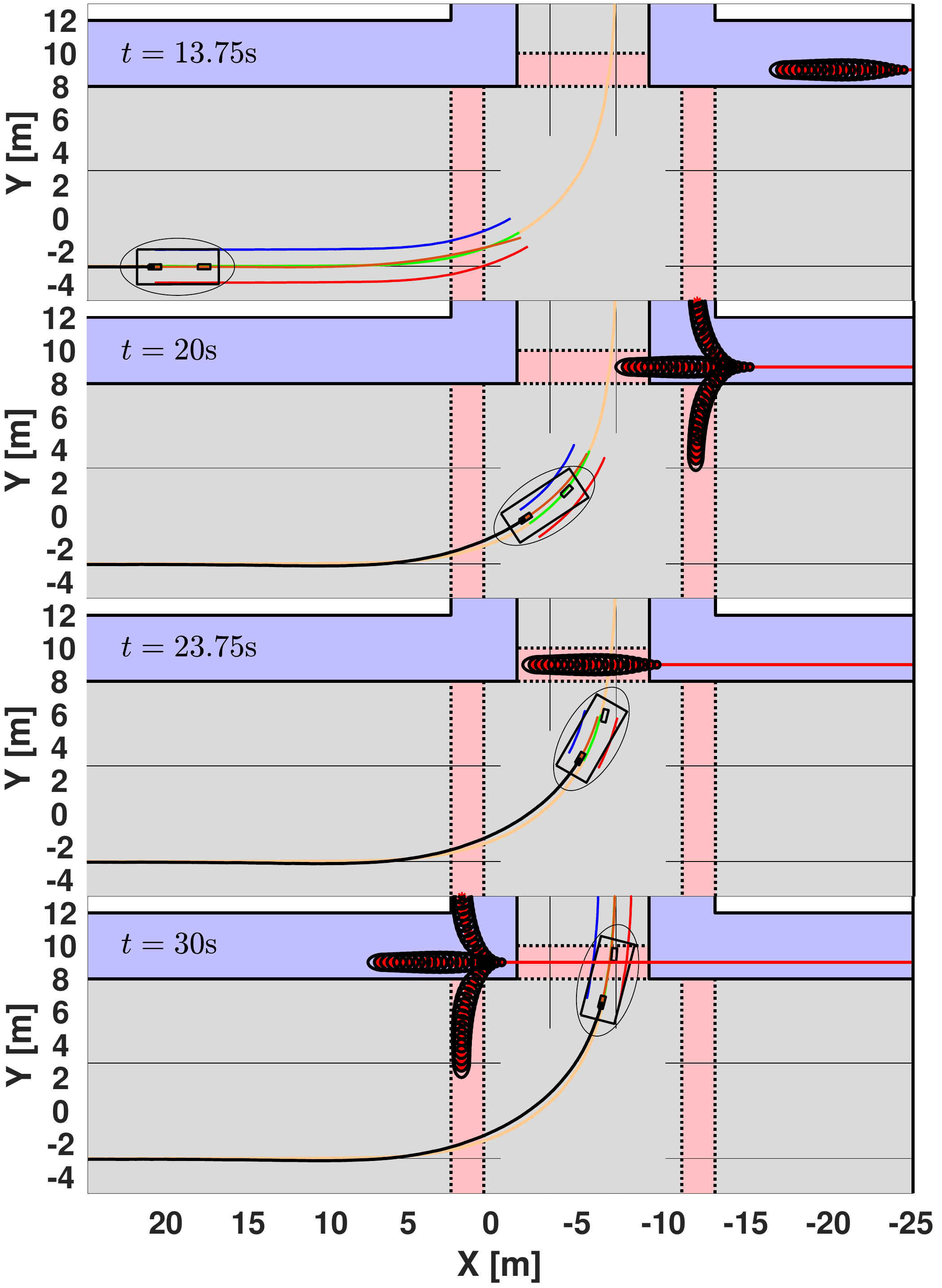}
		}}
		\caption{Open-loop solutions across different time instants. The color scheme matches the one of Fig.~\ref{fig:scenarioPedestrian2}.
		}
		\label{fig:scenarioPedestrian}
		\vspace{0em}
	\end{figure}

\begin{figure}[t!]
	\centering
	\mbox{\parbox{.48\textwidth}{
			\centering
			\includegraphics[width=0.85\linewidth]{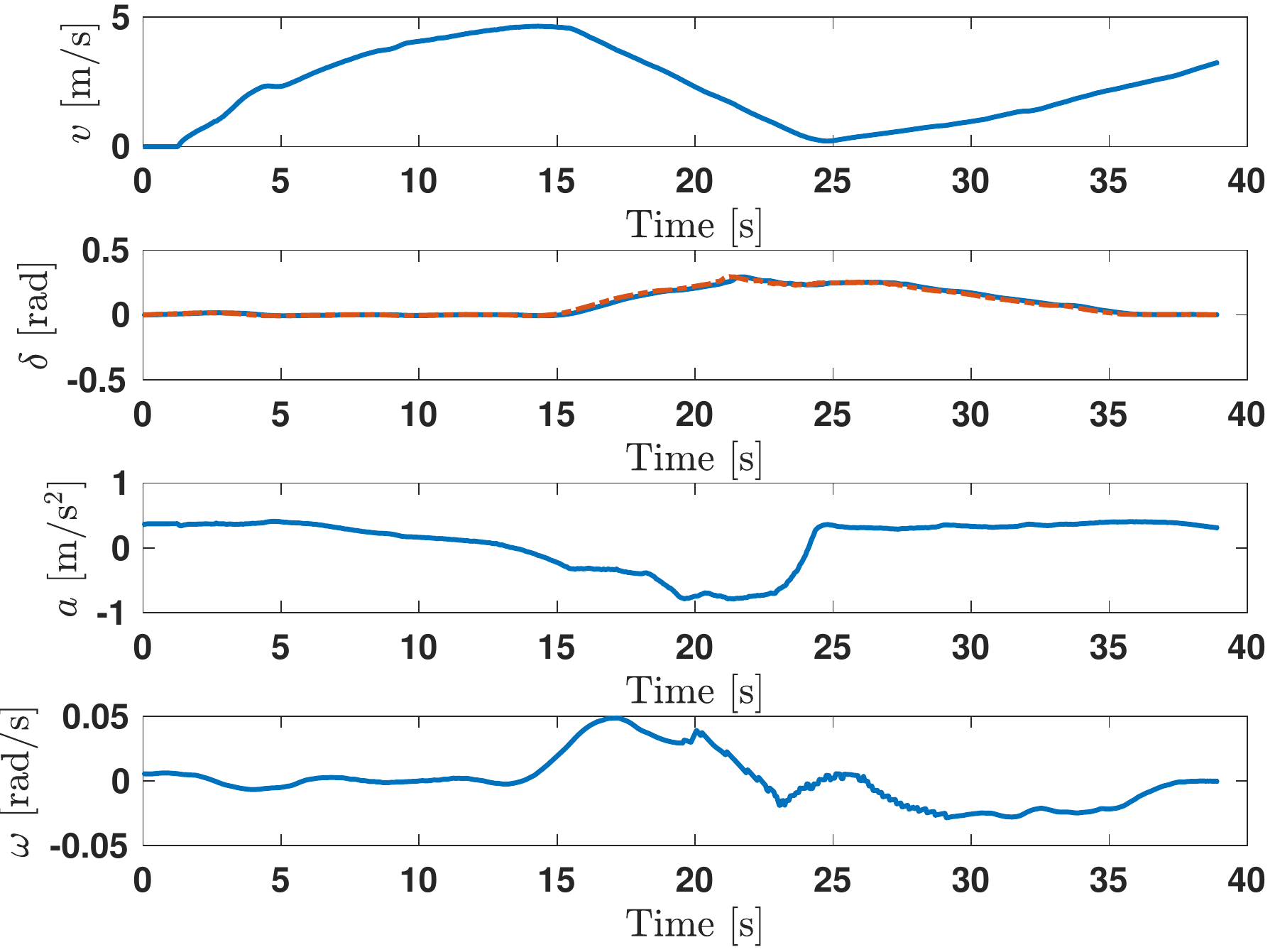}
	}}
	\caption{Closed loop states from Fig.~\ref{fig:scenarioPedestrian}. The plots match the same states as those of Fig.~\ref{fig:scenarioPedestrianStates2}. 
	}
	\label{fig:scenarioPedestrianStates}
	\vspace{-1.5em}
	\end{figure}

	Fig.~\ref{fig:scenarioPedestrian} shows the open-loop solutions across different time instants for the experiments at the test-track. The same controller from the simulation is used with the reference velocity $r_k^v=5\mathrm{m}/\mathrm{s}$. 
	For time $t=13.75\mathrm{s}$ the open-loop solution is not affected by the pedestrian constraints, and the controller is generating inputs only to stay on the path. For time $t=20\mathrm{s}$ the prediction algorithm propagates pedestrian positions through the intersection, where one of the predictions crosses the vehicle's path. Since the predicted positions are transformed into constraints, the vehicle is forced to slow down and make sure that the pedestrian passes. This can subsequently be seen in times $t=23.75\mathrm{s}$ through $t=30\mathrm{s}$. Fig.~\ref{fig:scenarioPedestrianStates} shows the corresponding  closed-loop states and control inputs. 
	The computer running the controller sent actuation signals at a rate of $20$Hz. The actual runtime of the framework, including the pedestrian prediction method, resulted in an average solution time of $10.3$ms, with a standard deviation of $2$ms, and longest timing of $19.7$ms. Thus, the runtime leaves a great margin to further increase the algorithmic complexity, or the sampling rate of the control scheme. In addition, since the controller ran in a Virtual Machine and not on the native computer operating system, there is a possibility to further reduce the computational time.
	
	\section{Conclusions}\label{sec:conclusions}
	In this paper we considered autonomous driving in urban environments with pedestrian crossings. We propose a general framework that solves the trajectory planning, and the longitudinal and lateral vehicle control problem both in simulations and real experiments. In addition, by combining predictions of the environment, it is shown that behaviors, such as slowing down, or stopping for a crossing pedestrian, are naturally included without any noteworthy increase in complexity. As shown, the framework already presents real-time performance without being highly optimized, except for the choice of sensitivity generation and QP solver. 
	
	Future work will aim to research situations where pedestrians suddenly appear to the framework due to sensor occlusion. In addition, we will aim to further verify the framework in more complex intersections with crossing pedestrians. Finally, future research will aim at developing a control scheme with recursive feasibility guarantees, which is robust to unexpected events.
	





	\bibliographystyle{IEEEtran} 
	\bibliography{references}

\end{document}